\documentclass[prc,showpacs,twocolumn]{revtex4}
\usepackage{amsmath}
\usepackage{graphicx}

\def\pt{$p_T$}
\def\dis{distribution}

\def\bq{\begin{equation}}
\def\eq{\end{equation}}

\begin{document}

\title{Relationship Between the Azimuthal Dependencies of Nuclear Modification Factor and Ridge Yield}
\author {Rudolph C. Hwa$^1$ and Lilin Zhu$^2$}
\affiliation
{$^1$Institute of Theoretical Science and Department of
Physics\\ University of Oregon, Eugene, OR 97403-5203, USA\\
$^2$Institute of Particle Physics, Hua-Zhong Normal
University, Wuhan 430079, P.\ R.\ China}

\begin{abstract} 
 
The azimuthal angular dependence of the nuclear modification factor $R_{AA}(p_T,\phi,N_{\rm part})$ recently obtained by PHENIX is related at low \pt\ to the trigger $\phi$ dependence of the ridge yield as measured by STAR in a framework in which the azimuthal anisotropy is driven by semihard scattering near the surface. Careful consideration of the initial geometry leads to the determination of a surface segment in which the production of semihard partons are responsible for the $\phi$ dependence of the inclusive \dis\ on the one hand, and for the angular correlation in ridge phenomenology on the other. With $v_2$ also being well reproduced along with $R_{AA}$ and ridge yield, all relevant $\phi$ dependencies in heavy-ion collisions can now be understood in a unified
description that emphasizes the ridge production whether or not a trigger is used.

\end{abstract}

\maketitle
\section{INTRODUCTION}

Recent measurement of $\pi^0$ production in heavy-ion collisions, expressed in terms of the nuclear modification factor $R_{AA}$, shows interesting dependence on the azimuthal angle $\phi$ at various centralities and transverse momenta $p_T$ \cite{saf}.  Those properties in $\phi$ are lost when the azimuthal anisotropy is summarized in terms of the elliptic flow coefficient $v_2$.  Similarly, the production of ridge in association with triggers has revealing behavior in $\phi_s$, the azimuthal angle of the trigger relative to reaction plane, at various centralities \cite{af}, but those properties are also lost upon integration over all $\phi_s$ in the determination of the total ridge yield \cite{jp, abe}.  In this paper we show that these two phenomena are related, even though $R_{AA}(p_T, \phi, N_{\rm part}$) is a measure of the single-particle distribution, while the ridge yield $Y^R(\phi_s, N_{\rm part})$ is a measure of the correlation between trigger and associated particles in specific $p_T$ ranges.

In \cite{saf} the attention is given mainly to hard processes at high $p_T$, but it is noted that at low $p_T$ and large $\Delta \phi$ around $\pi/2$ (the $\phi$ angle relative to the reaction plane) $R_{AA}$ is nearly constant in $N_{\rm part}$.  Such a curious property calls for an explanation.  A difference in the temperatures along the $x$- and $y$-directions in the transverse plane has been discussed in the Buda-Lund model \cite{ctl,ctc} where $v_2$ can be expressed in terms of a scaling variable that depends on the difference between those temperatures.  However, being a hydrodynamical study it does not consider explicitly the $\phi$ dependence.  Azimuthal anisotropy at low $p_T$ that does not depend on the assumption of rapid thermalization has been investigated in the context of semihard scattering and the ridges that are produced \cite{rh,chy}.  There also $v _2$ is derived without detailed consideration of the $\phi$ variable.  The azimuthal correlation between the ridge and trigger particles has, however, been studied in the correlated emission model (CEM) \cite{ch}.  It is from that study that we shall find in this paper an explanation of the $\phi$ dependence on  $R_{AA}$.

Since  $R_{AA}$ was introduced originally for the purpose of exhibiting the effect of jet quenching at high $p_T$, it is actually inappropriate to describe the behavior at low $p_T$.  In  the next section we shall introduce a modified $R_{AA}$ that is more suitable.  The present paper is focused exclusively on $p_T< 2$ GeV/c.  It is conventionally thought that at such low $p_T$ only soft physics is involved, a view that has been regarded as being too simplistic on account of the semihard scattering that can enhance the soft thermal partons \cite{rev,qgp}.  Whether or not trigger particles are used to select a subset of events, semihard partons are pervasive and generate ridges that affect both single-particle and dihadron distributions.  That is why we are able to relate the two azimuthal features of the hadronic observables that have been determined by the two RHIC experiments \cite{saf,af}.

The focus of this paper is on the azimuthal dependencies of $R_{AA}(p_T, \phi, N_{\rm part})$ and the ridge yield $Y^R(\phi_s, N_{\rm part})$ at midrapidity, where data exist \cite{saf,af}. We leave out completely any consideration about longitudinal correlation, on which there is experimental evidence that the ridge particles assoicated with a trigger can be widely separated in rapidity with $\Delta\eta\ ^>_\sim\ 4$ \cite{pho}, although autocorrelation without trigger shows a ridge with $\Delta\eta \approx 2$ \cite{ja}. The physics for longitudinal correlation is very different from that of azimuthal correlation at midrapidity. We regard them as separate issues to be treated separately, though all basic problems are related at some deep level that we do not address here.

\section{CONNECTION BETWEEN $R_{AA}(p_T, \phi, N_p)$ AND THE RIDGE}

The main content of $R_{AA}(p_T, \phi, N_p)$ is the single-particle distribution of  $AA$ collisions, which we abbreviate as $\rho _1$:
\begin{eqnarray}
\rho_1 (p_T,\phi, N_p ) = {dN_{AA} \over p_T \, dp_T \, d\phi}(N_p) \quad ,
\label{1}
\end{eqnarray}
where $N_p$ is a shortened notation for the number of participants, $N_{\rm part}$.  In the following we shall consider only the production of $\pi ^0$, since that is what is measured in \cite{saf} with high accuracy.  The properties of $\rho_1 (p_T,\phi, N_p )$ are usually presented, especially in hydrodynamical studies, in terms of the $p_T$ spectra (average of $\rho_1$ over $\phi$) and the elliptical flow coefficient $v _2 (p_T)$ (second harmonic in $\phi$).  The conceptual basis of flow deemphasizes the role of hard scattering among partons and offers a fluid description of the bulk medium at low $p_T$.  In that description azimuthal anisotropy in non-central collisions is regarded as the consequence of asymmetric pressure gradient in the transverse plane at early time, assuming fast thermalization.  However, that is not the only approach that can claim physical relevance, although it is the conventional view backed by a large number of investigations based on hydrodynamics \cite{pk2}.

An alternative view is to regard semihard scattering near the surface of the nuclear overlap as the driving force of the azimuthal anisotropy, since at low enough virtuality such scattering processes are pervasive throughout the medium, and when they occur near the surface of the almond-shaped initial configuration of the dense system, the semihard partons can emerge from the medium and not only hadronize as intermediate-$p_T$ jets, but also generate ridge particles at lower $p_T$ \cite{rev,qgp}.  Many issues are involved in the above statement.  First, the fact that semihard scattering cannot be calculated reliably in pQCD does not mean that its effects are not important.  Second, those effects are sensitive to the geometrical shape of the initial system at early time, since if the parton's transverse momentum is $k_T \sim $ 2-3 GeV/c, the time scale involved is $\ ^<_\sim\  0.1$ fm/c.  The semihard jets created near the surface give rise to non-trivial $\phi$ dependence of low-$p_T$ partonic distribution independent of the validity of the notion of pressure gradient at $\tau < 1$ fm/c \cite{rh}.  Third, those $\phi$-dependent soft partons hadronize and form the ridge structure observed by STAR whether with trigger \cite{af,jp,abe,jb} or without trigger \cite{ja}.  A theoretical description of the connection between ridges and $v_2$ in the approximation of using a very simple geometrical picture is given in \cite{rh,chy}.  It is along the same line of reasoning, but using a more realistic treatment of the initial geometry, that we relate in this paper all features of $\phi$-dependent observables, namely:  $R_{AA}(p_T, \phi, N_p)$, $v_2(p_T, N_p)$ and ridge yield $Y^R(\phi _s, N_p)$ for $p_T < 2$ GeV/c.

For hard scattering calculable in pQCD it has been shown that the Landau-Pomerachuk-Migdal interference effect suppresses the radiative energy loss of a hard parton in the initial phase of its trajectory \cite{bai, gw}. Applying the inference to semihard partons, despite uncertain validity, there can be an initial time interval when no energy loss may occur, the consequence of which is that the creation points of the relevant semihard partons can be farther away from the surface. What count for the ridge formation are those semihard partons that do lose energy on the way out and drive the azimuthal anisotropy. A shift in the time period when that occurs in the trajectory of a parton in the medium does not alter the relationship we propose between semihard scattering and ridge formation.

An important point to stress in distinguishing our approach from the conventional hydro approach is that the bulk of what is calculated in the latter contains $\phi$ dependence at $p_T < 2$ GeV/c and serves as the background of higher $p_T$ hard processes, while the bulk in our approach has no $\phi$ dependence and it is the ridge component at  $p_T < 2$ GeV/c that gives rise to the $\phi$ dependence of the soft component.  Thus we write
\begin{eqnarray}
\rho_1 (p_T,\phi, N_p ) = B  (p_T, N_p ) + R(p_T,\phi, N_p ) 
\label{2}
\end{eqnarray}
with $B$ and $R$ denoting bulk and ridge, respectively, for $p_T < 2$ GeV/c.  

At a fixed $p_T, \rho_1 (p_T,\phi, N_p )$ increases rapidly with $N_p$, but $R_{AA}(p_T, \phi, N_p)$ decreases with $N_p$ except when $\phi \approx \pi/2$.  The relationship between them is
\begin{eqnarray}
R_{AA} (p_T,\phi, N_p ) = Z (p_T,\phi, N_p ) \rho_1 (p_T,\phi, N_p ),
\label{3}
\end{eqnarray}
where the rescaling factor is
\begin{eqnarray}
 Z (p_T,\phi, N_p ) = \left[ N_c(N_p) dN_{pp}/p_T dp_T d\phi \right]^{-1},
\label{4}
\end{eqnarray}
$N_c$ being the number of binary collisions, short for $N_{\rm coll}$, and $dN_{pp}/p_T dp_T d\phi $ the single $\pi ^0$ inclusive distribution in $pp$ collisions. This rescaling factor is designed to give $R_{AA}$ a quantitative description of the effect of jet quenching at high $p_T$. But at low \pt, which is the region of interest to us in this paper, the more relevant rescaling factor is
\begin{eqnarray}
 Z' (p_T,\phi, N_p ) = \left[ N_p\, dN_{pp}/p_T dp_T d\phi \right]^{-1},
\label{5}
\end{eqnarray}
where $N_p$ is used instead of $N_c$ because of the dominance of soft processes.  Thus we define a different nuclear modification factor
\begin{eqnarray}
 R'_{AA}(p_T,\phi, N_p ) =  Z' (p_T,\phi, N_p )  \rho_1(p_T,\phi, N_p ).
\label{6}
\end{eqnarray}
From the data on $R_{AA}$ given in \cite{saf}, we can determine $R'_{AA}$.
We illustrate their differences by showing  in Fig.\ 1  (a) the original $R_{AA}(p_T,\phi, N_p )$ and (b) the modified $R'_{AA}(p_T,\phi, N_p )$, 
for $\phi$ in the ranges $0<\Delta\phi<15^{\circ}$ and $75^{\circ}<\Delta\phi<90^{\circ}$   for  $1.0 < p_T < 1.5$ GeV/c.
  In the following we shall for brevity use $\phi=0$ and $\pi/2$ to denote the two angular ranges, although in calculations that involve comparison with data we shall use $\phi=7.5^{\circ}$ and $82.5^{\circ}$, respectively.  The error bars in Fig.\ 1(b) are calculated from the errors on $N_p$ and $N_c$, neglecting the errors on $R_{AA}$ which are much smaller. 
  Comparison can also be made for  $1.5 < p_T < 2.0$ GeV/c, which we omit here since they are similar to those in Fig.\ 1.
  The effective values of the two \pt\ ranges will be 1.21 and 1.71 GeV/c given numerically in  tables in \cite{saf}.   
  We see in Fig.\ 1(a) that 
  $R_{AA}$ for $\phi=0$ decreases rapidly with $N_p$, but  $R'_{AA}$ in (b) is nearly independent of $N_p$ for $\phi = 0$.
  Whereas  $R_{AA}$ at $\phi=\pi/2$ is nearly constant,  $R'_{AA}(\phi=\pi/2)$
 increases linearly with $N_p$.  Intuitively, these properties of the $N_p$ dependence for $R'_{AA}$ are easier to understand, since the elliptical shape of the initial nuclear overlap is relatively independent of the impact parameter $b$ when viewed from the broad side at $\phi = 0$, but becomes narrower as $b$ increases, when viewed from $\phi = \pi/2$ where one sees mainly the narrow side of the ellipse.  Our problem is to reproduce these properties of  $R'_{AA}$ quantitatively in a framework where the decomposition of $\rho_1$ into the two components in Eq.\ (\ref{2}) plays an important role.

As described in \cite{rh,chy} and summarized in the beginning of this section, the azimuthal anisotropy arises from copious semihard scattering near the overlap surface; the produced semihard partons lose energy to the medium, whose enhanced thermal partons hadronize into the ridge particles that carry the footprint of the initial geometry.  The $\phi$ dependence of the ridge distribution $R(p_T,\phi, N_p )$ is therefore expected to be minimal at large $N_p$ (small $b$), but increases with decreasing $N_p$, thus opening the gap between $\phi = 0$ and $\pi/2$ as shown in Fig.\ 1.  The recoil semihard partons directed toward the interior of the dense medium are thermalized and retain no memory of the initial geometry by the time they are hadronized; together with all the other soft processes that occur in the interior they form the bulk that is $\phi$ independent.  With $B (p_T, N_p )$ in Eq.\ (\ref{2}) being independent of $\phi$, it is possible to calculate $v_2(p_T, N_p )$, whose value depends on the relative magnitudes of 
$B(p_T, N_p )$ and $R(p_T,\phi, N_p )$.  The absolute magnitude of $R(p_T,\phi, N_p )$ can be related to the ridge yield that depends on the trigger angle $\phi _s$ in dihadron correlation \cite{af}.  Thus we have a tightly constrained system that restricts the options available to finding a satisfactory physical basis to explain all the relevant experimental features.

\begin{figure}[tbph]
\hspace{-.3cm}
\includegraphics[width=0.45\textwidth]{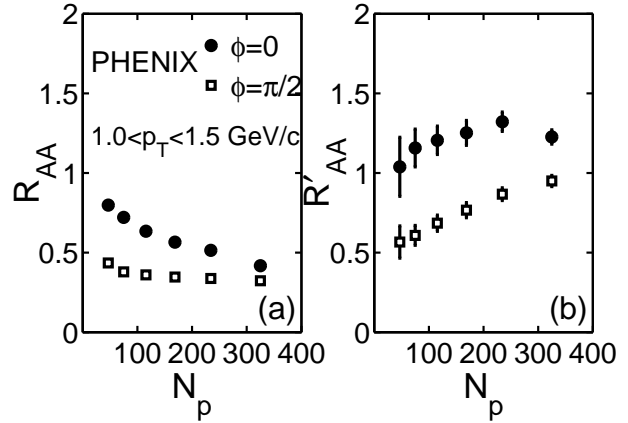}
\caption{(a) Original $R_{AA}$ and (b) modified $R'_{AA}$ at fixed \pt\ in the range $1.0<p_T<1.5$ GeV/c for $\phi$ in the ranges $0<\Delta\phi<15^{\circ}$, denoted by $\phi=0$ (in solid circle), and  $75^{\circ}<\Delta\phi<90^{\circ}$, denoted by $\phi=\pi/2$ (in open squares). The data are from \cite{saf}.}
\end{figure}

To summarize this section, we have described an overall picture in which the $\phi$ and $N_p$ dependencies of $R_{AA}(p_T,\phi,N_p)$ are to be related to those of the ridge structure at low $p_T$. In the following sections we implement this picture with concrete calculations.

\section{RIDGE AND DIHADRON CORRELATION}

The discussion in the preceding section is concerned mainly with the single-particle distribution $\rho_1 (p_T,\phi, N_p )$.  However, ridgeology (short for ridge phenomenology) involves two-particle correlation. Let us denote two-pion distribution after background subtraction by 
\begin{eqnarray}
\rho_2 (p_1, \phi_1, p_2, \phi_2 ) = {dN^{\pi \pi}_{AA} \over p_{1_T} dp_{1_T} d\phi _1 p_{2_T} dp_{2_T} d\phi _2}(N_p) ,
\label{7}
\end{eqnarray}
where the subscript $T$ will for brevity be omitted, since all momenta will hereafter be restricted to the transverse plane with $|\eta| < 0.35$.  We shall use label 1 to refer to the trigger  and label 2 to a particle in the ridge.  Thus, we have $\Delta \phi = \phi_2 - \phi_1$ for the correlation angle, not to be confused with the notation in \cite{saf} where $\Delta \phi$ denotes the angle of a pion relative to the reaction plane.  When the reaction plane is used as the reference for triggers in correlations (as will be done in the following), then $\phi_1$ is identified with $\phi_s$ in \cite{af}.  The dihadron correlation distribution has two components, which are referred to as Jet and Ridge \cite{jp,abe}, the former being the peak in $\Delta \eta$ that sits above the pedestal that is the latter with a wide spread in  $\Delta \eta$.  We write them without the arguments as 
\begin{eqnarray}
\rho_2 = \rho^J_2 + \rho^R_2 .
\label{8}
\end{eqnarray}
We shall consider only the latter, assuming that in ridgeology the former has been subtracted out from the dihadron correlation function.

The stage is now set for us to discuss the crucial point in our physical basis that relates $\rho^R_2 (p_1, \phi_1, p_2, \phi_2)$ to the ridge component $R(p_T, \phi, N_p)$ in Eq.\ (\ref{2}) for the single-particle distribution.  Usual ridgeology involves the use of a trigger and measures the momentum of another particle associated with it.  For $p_1$ in the range 3-4 GeV/c as studied by STAR in \cite{af}, the parton that produces the trigger is semihard, i.e., its transverse momentum is in the intermediate $k_T$ region, $< 5$ GeV/c.  PHENIX uses even lower trigger momentum ($2 < p_1< 3$ GeV/c) to study the away-side structure \cite{wh}.  The azimuthal properties of the ridge structure as functions of $\phi_1$ and $\phi_2$ separately provide significant insight into the correlation between the semihard and the soft thermal partons that does not depend sensitively on the trigger momentum \cite{af}.  A model on that azimuthal correlation has successfully reproduced the data on the $\phi_s$ dependence of the ridge yield \cite{ch}; moreover, a prediction on the asymmetry of the ridge structure in that model (CEM) has recently been verified by an analysis of the STAR data \cite{pn}.  The significance of that finding as related to our present problem is that the ridge yield studied in \cite{af,ch} involves an explicit integration of the $\Delta \phi$ distribution of $\rho^R_2$ over the ridge angle $\phi_2$, while the ridge component $R(p_T,\phi, N_p )$ in $\rho_1$ involves an implicit integration of the same over the trigger angle $\phi _1$.  A successful description of the correlation between $\phi_1$ and $\phi_2$ is therefore a crucial input for the determination of $R(p_T,\phi, N_p )$.

An important part of the physics implied in the above is that the direction of the semihard parton that generates the ridge need not be identified by a trigger particle, i.e., a ridge can exist whether or not there is a trigger.  A ridge particle at $\phi _2$ can be due to semihard partons at a range of $\phi _1$ around $\phi _2$, so for an inclusive ridge distribution in $\phi _2$, the value of 
$\phi _1$ should be integrated over, relative to every fixed $\phi _2$.  If we omit the arguments $p_T$ and $N_p$ in $R(p_T,\phi, N_p )$ and identify $\phi = \phi_2$, then we have
\begin{eqnarray}
R(\phi _2) \propto \int d\phi_1\ \rho^R_2 (\phi_1, \phi_2) ,
\label{9}
\end{eqnarray}
where the proportionality factor involves the integration over $p_1$, which will not be considered explicitly.  On the other hand, the ridge yield determined in \cite{af} integrates over all $\phi _2$ in the range $\left| \Delta \phi \right| < 1$, so we have
\begin{eqnarray}
Y^R (\phi _1) \propto \int d\phi_2\ \rho^R_2 (\phi_1, \phi_2)  ,
\label{10}
\end{eqnarray}
where the proportionality depends on the number of triggered events in prescribed ranges of $p_1$ and $p_2$, since $Y^R$ is the per-trigger yield.  The effective range of integration in both Eqs.\ (\ref{9}) and (\ref{10}) depends on the correlation width in $\rho^R_2 (\phi_1, \phi_2)$ and is much narrower than 1 rad.

It should now be clear why the $\phi$ dependence of $R_{AA}$ in the PHENIX data \cite{saf} should in our view be related to the ridgeological study by STAR \cite{af}.  There is, however, more complication than just the correlation in $\Delta \phi$.  The dependence on centrality that has not been made explicit in Eqs.\ (\ref{9}) and (\ref{10}) is evident in both \cite{saf} and \cite{af}.  To achieve the correct dependence on $N_p$ requires a careful consideration of the geometry of the problem, which is the subject of the next section.

A summary of this section is that the distribution of ridge particles and the ridge yield are two sides of the same coin, the former being the projection onto $\phi_2$ of the two-particle distribution $\rho_2^R(\phi_1,\phi_2)$, the latter being the projection of the same onto $\phi_1$.

\section{GEOMETRICAL CONSIDERATIONS}

The correlation between a semihard parton at $\phi _1$ and a ridge particle at $\phi _2$ has been studied in the CEM that involves several subprocesses \cite{ch}. (a) The semihard parton loses energy to the medium.  (b) Successive radiation by the parton enhances the thermal partons in the vicinity of the trajectory.  (c)  The enhanced thermal partons are carried by local flow to various points on the surface in directions normal to the surface.  (d)  Hadronization of those enhanced thermal partons by recombination generates the ridge particles that are centered around the average flow direction.  (e)  Density of the ridge is highest when the flow directions are in alignment with the direction of the semihard parton.  (f)  The resultant correlation between $\phi_1$ and $\phi_2$ can be described by the convolution of a Gaussian distribution in $\phi_1 - \psi$ (where $\psi$ is the local flow direction) with another distribution describing the fluctuation of $\phi _2$ from $\psi$.  In the midst of all those subprocesses that cannot be calculated from first principles, there are considerations of nuclear density, points of semihard scattering, emission points along the trajectories, and the exit points at the surface, all of which have been taken into account to the extent possible.  What we can adapt from that study in ridgeology is the central result that there is a correlation
\begin{eqnarray}
C(x, y, \phi_1) = \exp \left\{- {\left[ \phi_1 - \psi (x, y) \right]^2 \over 2 \sigma ^2} \right\}
\label{11}
\end{eqnarray}
between the semihard parton direction $\phi_1$ and the flow direction $\psi(x, y)$ at the exit point $(x, y)$ on the surface, with $\sigma = 0.33$.  The angle between the ridge particle $\phi_2$ and $\psi$ due to fluctuations has negligible effect on the results.

For every semihard parton direction $\phi_1$ a variation in the starting point of the trajectory in the nuclear overlap leads to a variation in $\psi(x, y)$.  This important property can be turned around and applied to the ridge component of the single-particle distribution.   With $\phi_2$ identified as $\psi$, for every ridge particle detected at $\phi_2$ there is a variety of semihard partons in direction $\phi_1$ that can contribute to that ridge depending on where the exit point $(x, y)$ is, so long as $|\phi_1 - \phi_2|$ is within the Gaussian width of Eq.\ (\ref{11}).  The point to stress here is that for $R(p_T,\phi, N_p ) $ in the single-particle distribution in Eq.\ (\ref{2}), $\phi$ is $\phi _2$ and the semihard parton at $\phi_1$ is undetected and should be integrated over.

The scheme described above can be implemented by concrete calculation because the main input is angular correlation using early time geometry.  Hadronization, of course, occurs at late time, and we do not rely on hydrodynamics to describe the evolution of the system in space-time.  Our assumption is that the ridge particles in the final state have the same $\phi$ distribution as that of the enhanced thermal partons at early time, which in turn is prescribed by Eq.\ (\ref{11}).   The bulk component that has no preferred direction to expand (not being the same as the hydrodynamical bulk) is isotopic, and becomes the background of the ridge that carries all the information about the anisotropy of the initial geometry.

When two nuclei of radius $R_A$ collide at impact parameter $b$, the almond-shaped overlap has width and height given, respectively, by 
\begin{eqnarray}
w = 1 - b/2, \qquad h = (1-b^2/4)^{1/2},
\label{12}
\end{eqnarray}
where all lengths are normalized by $R_A$.  The appropriate geometry that can best describe the intial configuration almost immediately after impact is the ellipse
\begin{eqnarray}
({x \over w} )^2 + ({y \over h} )^2 = 1 \ .
\label{13}
\end{eqnarray}
The flow direction at the surface is normal to that surface, so its azimuthal angle is 
\begin{eqnarray}
\psi (x, y) = \tan^{-1} \left({w^2 y \over h^2x}\right) \ .
\label{14}
\end{eqnarray}
It should be noted that although we refer to $\psi$ as the flow direction, the use of those words is mainly a compact way to name the direction normal to the surface, as defined in Eq.\ (\ref{14}). We place no emphasis on hydrodynamics as the basis for that description, especially in the initial phase of the evolution of the medium.
The implication of Eq.\ (\ref{11}) is that for a fixed semihard parton at $\phi_1$, the ridge particle at $\phi_2$ identified with $\psi (x, y) $ can deviate from $\phi_1$ at most by $\sigma$, as illustrated in Fig.\ 2(a).  By the same reasoning, for a fixed $\phi_2$ the possible semihard partons that can contribute to the corresponding ridge must have their $\phi_1$ that can deviate from $\phi_2$ by no more than $\sigma$, as illustrated in Fig.\ 2(b).  That means the exit point $(x, y) $ of the semihard parton at the surface is restricted to a certain range, which we shall denote by $S(\phi_2, b)$.

\begin{figure}[tbph]
\hspace{1cm}
\includegraphics[width=0.4\textwidth]{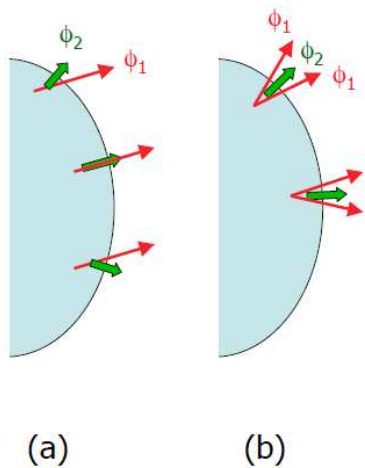}
\caption{(Color online) Sketches to illustrate the relationship between the semihard parton's direction $\phi_1$ (in thin red arrows) and the ridge particle's direction $\phi_2$ (in thick green arrows). (a) Three possible semihard partons all having the same $\phi_1$, but generating ridge particles in different directions $\phi_2$ not deviating by more than $\sigma$ from $\phi_1$. (b) For a fixed $\phi_2$ (two being illustrated), $\phi_1$ of contributing partons are restricted to within a cone of width $\sigma$ around $\phi_2$.}
\end{figure}

From the geometry of the ellipse we can use the constraint of Eq.\ (\ref{13}) to write $x = w \cos \theta$ and $y = h \sin \theta$ so that $S(\phi_2, b)$ can be determined by
\begin{eqnarray}
S(\phi, b) &=& \int_{\rm arc} d \ell = \int \left[(dx)^2 + (dy)^2 \right]^{1/2}  \nonumber  \\
&=& \int ^{\theta_2}_{\theta _1} \left[ w^2 \sin^2 \theta + h^2 \cos^2 \theta\right]^{1/2} d \theta                \nonumber \\
&=& h \left[E(\theta_2, \alpha) - E(\theta_1, \alpha) \right] ,
\label{15}
\end{eqnarray}
where $E(\vartheta, \alpha)$ is the elliptic integral of the second kind
\begin{eqnarray}
E(\vartheta, \alpha)=\int ^{\vartheta}_0 (1 - \alpha \sin^2 \theta)^{1/2} d \theta .
\label{16}
\end{eqnarray}
In Eq.\ (\ref{15}) we have
\begin{eqnarray}
\alpha (b) &=& 1 - w^2/h^2 \ ,  \nonumber \\
\theta_i&=&\tan^{-1} \left({h\over w}\tan \phi_i\right), \  i=1,2, \  {\rm if} \  \phi_i\leq\pi/2 \ ,   \nonumber \\ 
\phi_1 &=& \phi - \sigma , \quad  \phi_2 = \phi + \sigma. \label{17}
\end{eqnarray}
If $\phi_2>\pi/2$, then 
\begin{eqnarray}
\theta_2={\pi\over 2}+\left|\cot^{-1} \left({h\over w}\tan \phi_2\right)\right|. \label{17a}
\end{eqnarray}Thus the physical meaning of $S(\phi, b)$ is that it is the segment of the surface through which the semihard parton can be emitted to contribute to a ridge particle at $\phi$. In Fig.\ 3 we show 
$S(\phi, b)$ for $\phi = 0$ and $\pi /2$, and $\sigma = 0.33$.  
The two curves start out with the same value at $b=0$, but develop a wide gap between them, as $b$ increases.
If they are plotted against $N_p$, then the behavior of $S(\phi, N_p)$ appear similar to the $N_p$ dependence of $R'_{AA}$ shown in Fig.\ 1(b) for $b<1.5$, corresponding to $N_p>50$, apart from a common background for the two sets of data points.  That is a strong hint that we are on the right track.

\begin{figure}[tbph]
\hspace{-.3cm}
\includegraphics[width=0.35\textwidth]{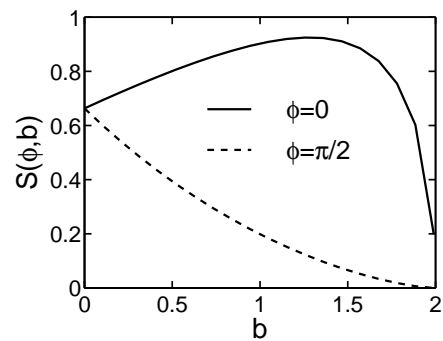}
\caption{Surface segment $S(\phi,b)$ vs normalized impact parameter $b$ for $\phi=0$ (solid line) and $\phi=\pi/2$ (dashed line) and $\sigma=0.33$.}
\end{figure}

Note that the initial slope of $S(0, b)$ at $b = 0$ is positive, but that of $S(\pi /2, b)$ is negative.  The latter can intuitively be understood because the ellipse is narrower than the circle when viewed from the top, so the segment of the surface within the width $\sigma$ decreases with increasing $b$.  In the former case the ellipse is flatter than the circle   when viewed from $\phi = 0$, so at fixed $\sigma$ the segment covers more of the surface as $b$ is increased near $b = 0$.  What is remarkable is that such detailed geometrical properties seem to be exhibited by the behavior of $R'_{AA}$ at high $N_p$.

It should be mentioned that our consideration so far has been exclusively in the transverse plane at midrapidity so that the use of elliptic geometry has been simple and adequate for the calculation of the surface segment $S(\phi,b)$ through which semihard partons are emitted at small $|y|$. Since semihard scattering occurs at early time ($\tau<0.2$ fm/c), the appropriate geometry is the overlap in the transverse plane of two Lorentz-contracted discs. Longitudinal geometry enters the problem in the consideration of local density later in this section.

We now assert as the basic assumption about the azimuthal anisotropy of the single particle distribution $\rho_1 (p_T, \phi, b)$ that it arises from the $\phi$ dependence of the ridge component that is proportional to $S(\phi, b)$.  That component depends on all three variables $p_T$, $\phi$ and $b$.  Since the thermal partons are exponential in $p_T$, we write in the factorizable form
\begin{eqnarray}
R (p_T, \phi, b) = e ^{-E_T (p_T) / T'} A(b)  S(\phi, b) \ , 
\label{18}
\end{eqnarray}
where the use of the transverse kinetic energy $E_T$ is discussed in \cite{chy} as a means to account for the hadronic mass $m_0$ with $E_T (p_T) = (p^2_T + m^2_0 )^{1/2}- m_0$.  The factor $A(b)$ depends only on $b$, since the other two factors in Eq.\ (\ref{18}) exhaust the $p_T$ and $\phi$ dependencies.  The bulk term in Eq.\  (\ref{2}) has no $\phi$ dependence, so we write as in \cite{chy}
\begin{eqnarray}
B(p_T, b) = {C^2 (b) \over 6} e ^{-E_T (p_T) / T} \ ,
\label{19}
\end{eqnarray}
where the prefactor $C^2/6$ follows from the thermal parton distribution $q^0 dN^B/dq_T d\phi = Cq_T \exp (-q_T/T)$ through recombination.  $C(b)$ is known from a previous study and is described as a power law in $N_p$ \cite{yt}, on which we shall make a slight modification below to render a better description of the data.  Since the ridge particles are due to the hadronization of the enhanced thermal partons, its $T'$ is higher than the bulk $T$ by a small amount discussed previously \cite{jp,chy} and will be detailed below.

The factorized form for $R(p_T, \phi, b)$ in Eq.\ (\ref{18}) is an approximation of a complicated expression that involves an integration over all points of semihard scattering in the nuclear overlap region and another integration along the trajectory of each semihard parton in the medium \cite{ch}, as well as over the angle $\phi _1$ in Eq.\ (\ref{9}).  The probabilities of the creation of a semihard parton and of its energy loss depend on the local density of the medium. We simplify all that by writing $A(b)$ in Eq.\ (\ref{18})  as being proportional to the effective density $D (x', y', b)$ at a representative point $(x', y')$ near the surface where semihard scattering takes place.  Factoring out $C^2 (b)/6$ explicitly which carries the dimension (GeV)$^{-2}$ for both $B(p_T,b)$ and $R(p_T, \phi, b)$, we write
\begin{eqnarray}
A(b) = {C^2 (b)\over 6} a D (x', y', b) \quad ,
\label{20}
\end{eqnarray}
where $a$ is a free parameter to be fixed by the magnitude of $v_2 (p_T, b)$ that determines the degree of azimuthal anisotropy.  As we have noted earlier, this is a highly constrained problem in which many pieces of data are brought to bear on the phenomenology.

For the local density $D (x', y', b)$ in the transverse plane we use the Glauber model to compute it, which, for an arbitrary point $(x, y)$, is, apart from an unspecified overall normalization, 
\begin{eqnarray}
D (x, y, b) =  L_A (x, y) \left[1 - e^{-\omega L_B(x, y)}  \right]  \nonumber \\
+  L_B (x, y) \left[ 1 - e^{-\omega L_A(x, y)} \right]\ ,   
\label{21}
\end{eqnarray}
where $\omega = 4.6$ for Au+Au collision \cite{ch} and $L_{A,B} (x, y) $ are the longitudinal lengths
\begin{eqnarray}
L_{A,B} (x, y) = {1\over \rho_0}\int_{-z_{A,B}}^{z_{A,B}} dz \rho(s,z) \ ,
\label{22}
\end{eqnarray}
and $\rho(r)$ is the nuclear density in Woods-Saxon form
\begin{eqnarray}
\rho(r) = \rho_0 \left[ 1+e^{(r-r_0)/\xi}\right]^{-1} 
\label{23}
\end{eqnarray}
with $r_0 = 0.92$, $\xi = 0.08 $, and $\rho _0 = 0.285$ where all lengths are in units of $R_A = 7$ fm and $\rho$ is normalized by $A$.  In Eq.\ (\ref{22})  $z_A^2=1-s^2,\   z_B^2=1-|\vec s-\vec b|^2$, and 
\begin{eqnarray}
s^2 = (x + b/2)^2 + y^2 \ .
\label{24}
\end{eqnarray}
The approximation we make to avoid averaging over all points $(x, y)$ in the initial overlap is to choose $(x', y')$ to be at a  representative distance $\delta r$ from the surface on the short axis of the almond.  Detailed study has shown that the layer in the medium just below the surface in which semihard partons are predominantly produced, leading to the ridge particles, is of thickness 0.2 \cite{ch}.  Thus we adopt $\delta r = 0.17$ for the average and choose 
\begin{eqnarray}
x' = 1 - b/2 - \delta r, \qquad y' = 0 .
\label{25}
\end{eqnarray}
In this way $D (x', y', b)$ is a calculable quantity for every $b$.  Its normalization is absorbed into the unknown parameter $a$ in Eq.\ (\ref{20}).

The treatment of the geometry of the collision problem in this section is far more realistic than the $\Theta (\phi)$ function used in \cite{rh,chy}.  The incorporation of the correlation properties found in CEM \cite{ch} in the determination of $S (\phi, b)$ endows the ridge component $R(p_T, \phi, b)$ with a reliably calculable $\phi$ dependence that can be tested as the physical origin of the anisotropy of the single-particle distribution $\rho _1  (p_T, \phi, b)$.

A way to summarize what has been done in this section is to use Fig.\ 2(b) and state that, for every ridge particle at $\phi_2$ indicated by the thick green arrow, the possible angles $\phi_1$ of the semihard partons that can contribute to it are within the cone shown by the thin red arrows, which subtends a segment on the elliptical boundary. That segment is quantified by $S(\phi,b)$ that we have calculated. It gives the $\phi$ dependence of the ridge \dis\ $R(p_T,\phi,b)$, whose $b$ dependence is affected by the medium density that is also calculable from geometry.

\section{ELLIPTIC FLOW COEFFICIENT $v_2 (p_T, b)$}

It is now straightforward to calculate $v_2$, defined by the second harmonic of $\rho _1$
\begin{eqnarray}
v_2 (p_T, b) = \left< \cos 2 \phi \right> = {\int ^{2 \pi}_0 d\phi \cos 2\phi \rho_1 (p_T, \phi, b)  \over  \int ^{2 \pi}_0 d\phi \rho_1 (p_T, \phi, b)  }\ .
\label{26}
\end{eqnarray}
Using Eqs.\ (\ref{2}) and (\ref{18}) we obtain
\begin{eqnarray}
v _2 (p_T, b)  = {\int ^{\pi/2}_0 d\phi \cos 2\phi S (\phi , b)  \over  K(p_T, b) + \int ^{\pi/2}_0 d\phi  S (\phi , b)}
\label{27}
\end{eqnarray}
where, owing to Eqs.\ (\ref{19}) and (\ref{20}), 
\begin{eqnarray}
K(p_T, b)   &=& {\pi B (p_T, b)  \over  2e^{-E_T (p_T)/T'}A(b)}  \nonumber \\
&=& {\pi \over 2aD(x', y', b)}  e^{-E_T (p_T)/T''}
\label{28}
\end{eqnarray}
with
\begin{eqnarray}
{1 \over T''}= {1 \over T} - {1 \over T'} \ .
\label{29}
\end{eqnarray}
Note that $C^2(b)$ in $A(b)$ and $B(p_T, b)$ are cancelled, leaving only the density $D(x', y', b)$ to prescribe the $b$ dependence in $K(p_T, b)$.

The integrals in Eq.\ (\ref{27}) are easy to evaluate, but the result on $v_2$ depends on the magnitude of $K(p_T, b)$, which has an undetermined parameter $a$.  Moreover, the suitable values for the inverse slopes $T$ and $T'$ require some discussion of the thermal distributions.  In the recombination model for heavy-ion collisions \cite{hy} the shower parton distribution S can be calculated from semihard and hard scattering, but the thermal parton distribution T is determined by phenomenology.  It is the hadron distribution at low $p_T$ that is fitted by an exponential form, from which is inferred the T  distribution through TT recombination in the case of meson. TS recombination is then calculated to obtain the effect of semihard scattering at intermediate $p_T$ .  What is done in \cite{hy} is to extract the inverse slope from the inclusive spectrum averaged over all $\phi$.  Ridgeology in \cite{rh,chy} separates $T'$ from $T$, but is based on a  very simple model on $\phi$ dependence.  In our more realistic study here the complication of $R(p_T, \phi, b)$ in Eq.\ (\ref{18}) cannot be isolated from the background $B(p_T,b)$ in Eq.\ (\ref{2}) without the knowledge of the parameter $a$ in Eq.\ (\ref{20}), which in turn cannot be determined separately from $T$ and $T'$.  An iterative process may be necessary, the first step of which is what we do now by taking $T'$ to be as given in \cite{hy}:  $T' = 0.317$ GeV, and $\Delta T = T' - T = 0.045$ GeV as given in \cite{jp}.  From Eq.\ (\ref{29})  follows $T'' = 1.916$ GeV.  With these values we can calculate $K (p_T, b)$ and then $v_2 (p_T, b)$ in Eq.\ (\ref{27}) with a suitable choice of $a$.

\begin{figure}[tbph]
\hspace{-.3cm}
\includegraphics[width=0.4\textwidth]{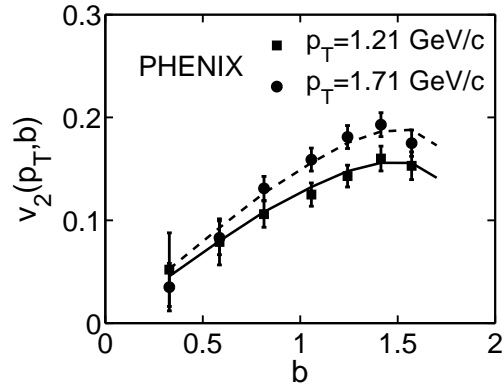}
\caption{Elliptic flow coefficient $v_2(p_T,b)$ vs $b$ for $p_T=1.21$ GeV/c (solid) and 1.71 GeV/c (dashed). Data are from \cite{saf}.}
\end{figure}

In Fig.\ 4 we show our calculated results for $v_2 (p_T, b)$ for $p_T = 1.21$ (solid line) and 1.71 GeV/c (dashed line) exhibiting excellent agreement with the data that have the corresponding $\left<p_T \right>$ \cite{saf}.  The one parameter we adjust to get the best fit is $a = 0.47$.  The nature of the $b$ dependence is critically dependent on the properties of $S(\phi, b)$, while $K(p_T, b)$ plays a less sensitive role, being of order 1 for most values of $b$ at $p_T < 2$ GeV/c.  $K(p_T, b)$ is a measure of the relative strength of the bulk component $B(p_T, b)$ to the ridge component $R(p_T, \phi, b)$ in Eq.\ (\ref{2}) for single-particle distribution.  The fact that we have a good reproduction of the data on $v_2 (p_T,b)$ validates our approach to assigning the $\phi$ dependence of $\rho_1 (p_T, \phi, b)$ entirely to  $R(p_T, \phi, b)$.  That validation is strengthened when the same formalism can be shown to reproduce the data on the ridge yield in $\phi_s$-dependent dihardron correlation.  

To summarize, we have in this section made the first phenomenological demonstration that the $\phi$ dependence calculated in Sec. IV leads to a good description of the data on $v_2(p_T,b)$ using one free parameter $a$ for the overall magnitude of the azimuthal anisotropy.

\section{DEPENDENCE OF RIDGE YIELD ON TRIGGER DIRECTION}

We now consider the correlation between a trigger and the ridge particle associated with it.  It may be helpful to recapitulate what we have done.  For single-particle distribution the ridge component $R(p_T, \phi, b)$ is related to the dihadron correlation distribution by Eq.\ (\ref{9}), where $\rho^R_2$ is the ridge component of $\rho_2$ expressed in Eq.\ (\ref{8}).  Dominance of $\rho^R_2$ by the correlation function $C(\phi_1, \phi_2)$ shown in Eq.\ (\ref{11}) with $\phi_2 = \psi(x,y)$ leads to the study of the surface factor $S(\phi, b)$ given by  
Eq.\ (\ref{15}).  The integral in Eq.\ (\ref{9}) is carried out in Eq.\ (\ref{15}) that results in $R(\phi_2, b)\propto S (\phi_2, b)$, as shown explicitly in Eq.\ (\ref{18}).  Now, with the same $\rho^R_2 (\phi_1, \phi_2)$ we can calculate the ridge yield $Y^R(\phi_1)$ as given in Eq.\ (\ref{10}), where the integration of  $\rho^R_2 (\phi_1, \phi_2)$, dominated by $C(\phi_1, \phi_2)$, over $\phi_2$ results in 
\begin{eqnarray}
Y^R(\phi_1, b) \propto S (\phi_1, b) \quad . 
\label{30}
\end{eqnarray}
Thus what we know already about $S (\phi, b)$, shown in Fig.\ 3, need only be replotted to be compared to data.  

In \cite{af} the per-trigger ridge yield is analyzed for $p^{\rm trig}_T$ and $p^{\rm assoc}_T$ in the narrow ranges $3 < p^{\rm trig}_T <4$ GeV/c and  $1.5 < p^{\rm assoc}_T <2.0$ GeV/c in 6 bins of $\phi _s$ (trigger direction relative to the reaction plane) between $0$ and $\pi/2$ for two centralities 0-5 \% and 20-60 \%.   For the corresponding $b$ values we use $\bar b = 0.33$ for the former and average over $1.06, 1.24, 1.41$, and $1.57$ for the latter.  From $S(\phi,b)$ in Eq.\ (\ref{15}) we obtain the results for $Y^R (\phi, \bar b)$, shown by the solid lines in Fig.\ 5(a) and (b), compared to the data from \cite{af}.  The normalization of the theoretical curves is adjusted to render a good overall fit in Fig.\ 5(a), since the data on yield depend on the experimental cuts so their magnitudes are not predictable.  However, once the normalization factor for Fig.\ 5(a) is determined, that for Fig.\ 5(b) is not readjustable.  Our results show good agreement with the data both in the two $\phi_s$ dependencies and in the normalization for the non-central collisions. 

\begin{figure}[tbph]
\hspace{-.3cm}
\includegraphics[width=0.45\textwidth]{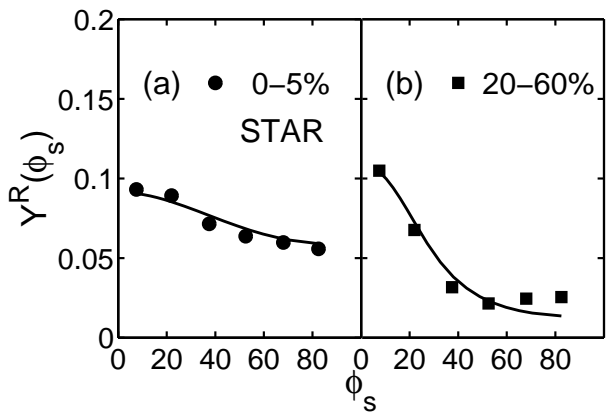}
\caption{Ridge yield per trigger vs trigger angle $\phi_s$ relative to reaction plane for (a) 0-5\% and (b) 20-60\% centralities. Solid lines are calculated results from Eq.\ (\ref{30}), and the data are from \cite{af}.}
\end{figure}

The mild decrease of $Y^R (\phi_s)$ with $\phi_s$ in Fig.\ 5(a) can be seen in 
 Fig.\ 3 when $b$ is restricted to $b = 0.33$.  The precipitous decrease in Fig.\ 5(b) corresponds to the large gap between the two curves of $S(\phi, b)$ for $\phi = 0$ and $\pi /2$ when $b$ is between $1.06$ and $1.57$.  Of particular interest is the increase of $S(\phi = 0, b)$ when $b$ is increased from $0.33$ to about $1.5$. Indeed, the data of $Y^R (\phi_s, b)$ in Fig.\ 5 show that the yield at the lowest $\phi_s$ is larger for 20-60\% than for 0-5\%, unlike the situation for all higher $\phi_s$ bins.  That is a remarkable confirmation of the validity of Eq.\ (\ref{30}) and of the properties of 
$S(\phi, b)$ that we have obtained.

Thus in this section we have shown the other side of the coin in the metaphor used at the end of Sec.\ III. The ridge yield is also well described by $S(\phi,b)$.

\section{NUCLEAR MODIFICATION FACTOR}

Having shown the existence of substantial support for the relevance of $S(\phi, b)$ to $v_2 (p_T, b)$ and $Y^R (\phi_s, b)$, we now return to $R_{AA}(p_T,\phi, b )$, the new data on which stimulated our present line of investigation in the first place.  Recall that in Sec.\ 2 an alternate nuclear modification factor  (NMF) $ R'_{AA}(p_T,\phi, N_p )$ is defined by rescaling with $N_p$ instead of $N_c$ for $p_T < 2$ GeV/c; its dependence on $N_p$ is shown in Fig.\ 1(b).  The way in which the data points for $\phi = 0$ and $\pi/2$ converge as  $N_p$ approaches 400 is very similar to the way that $S(0, b)$ and $S(\pi/2, b)$ converge as $b \to 0$ in Fig.\ 3.  The $\phi$-dependent part is, however, only the ridge component in $\rho_1 (p_T, \phi, N_p)$; the bulk component that is roughly of the same magnitude is the other piece yet to be added, rendering $ R'_{AA}(p_T,\phi, N_p )$ to behave as in Fig.\ 1.

The normalizations of both $B(p_T, N_p)$ and $R(p_T, \phi, N_p)$ in Eq.\ (\ref{2}) should now be determined.  Their relative magnitude has already been examined in connection with $K(p_T, b)$ in Eq.\ (\ref{28}).  Using Eqs.\ (\ref{18}), (\ref{19}) and (\ref{20}), let us re-express their sum as follows
\begin{eqnarray}
\rho_1 (p_T, \phi, N_p) = C^2 (N_p)\ f(p_T,\phi,N_p) \ ,   \label{31}
\end{eqnarray}
where
\begin{eqnarray}
f(p_T,\phi,N_p)={1\over 6}\left[e^{-E_T/T} + a D(b) S(\phi, b)e ^{-E_T/T'}\right].
\label{31a}
\end{eqnarray}
 The strength of the thermal partons in the medium, characterized by $C(N_p)$, is not calculable in our approach and has always been determined by fitting the low-$p_T$ hadronic data \cite{hy}.  Its dependence on $N_p$ has been quantified in \cite{yt} based on older data.  We now give a better determination of it in light of the newer data  that include also the $\phi$ dependence \cite{saf}.

It has been an assumption in our approach that the $\phi$ dependence of $\rho_1(p_T, \phi, N_p)$ is borne by $R(p_T, \phi, N_p)$ alone in Eq.\ (\ref{2}), and not by $B(p_T, N_p)$ at all. While that assumption has been tested indirectly by $v_2(p_T, b)$ in Fig.\ 4, we now confront it directly by recognizing that if it is valid, then $f(p_T, \phi, N_p)$ in Eq.\ (\ref{31a}), which we can calculate, must correctly describe the experimental $\phi$ dependence of $\rho_1(p_T, \phi, N_p)$, apart from an overall normalization. In other words, $C(N_p)$ in Eq.\ (\ref{31}) must be independent of $\phi$, a property that could not have been checked prior to any knowledge about the $\phi$ dependence of $\rho_1(p_T, \phi, N_p)$. With the data on the latter being now given in \cite{saf}, we have checked that $C(N_p)$ is indeed insensitive to $\phi$. We exhibit that finding in the following more revealing way.

\begin{figure}[tbph]
\hspace{-.3cm}
\includegraphics[width=0.4\textwidth]{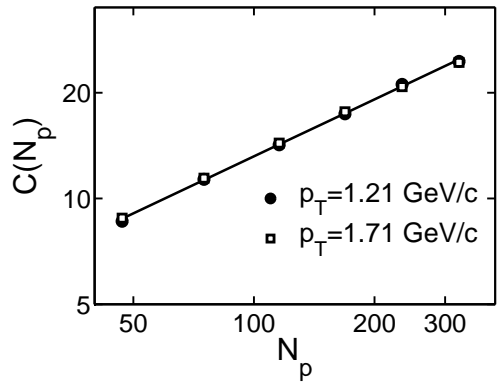}
\caption{The points are empirical $C(N_p)$ described in the text for \pt=1.21 GeV/c (solid circles) and 1.71  GeV/c (open squares). The straight line is a power-law fit.}
\end{figure}

If we use the experimental values for $\rho _1$ on the LHS of Eq.\ (\ref{31}) and the calculated values of $f(p_T, \phi, N_p)$ on the RHS, we can determine their averages over $\phi$ separately; we then take the square root of their ratio and regard the result as the empirical $C(N_p)$. Strictly speaking, it is possible for $C(N_p)$ thus extracted to depend on whether \pt\ is 1.21 or 1.71 GeV/c.
They are, however, shown to be independent of \pt\  in Fig.\ 6 by points depicted in different symbols that essentially overlap. We can fit them very well by a simple power law
\begin{eqnarray}
C(N_p) = 1.1\ N^{0.54}_p\ {\rm GeV}^{-1}\ ,
\label{32}
\end{eqnarray}
as shown by the straight line.
We can then use this formula in Eq.\ (\ref{31}) and exhibit the $\phi$ dependence of $\rho_1(p_T, \phi, N_p)$ in two ways.
First, we plot the experimental data of $\rho_1(p_T, \phi, N_p)$ vs $C^2(N_p)$ as points in Fig.\ 7 at the six values of $N_p$ treated as parametric variables and for  (a) $1.0<p_T<1.5$ GeV/c and (b) $1.5<p_T<2.0$ GeV/c. In each panel there are three sectors of $\phi$ values corresponding to $\Delta\phi=(0,15^{\circ}), (30^{\circ},45^{\circ})\ {\rm and}\ (75^{\circ},90^{\circ})$ analyzed in \cite{saf}, although the legend in the figure approximates them as integral fractions of $\pi$. The curves are the results of our evaluation of $f(p_T, \phi, N_p)$ at the corresponding values of the parameters. The agreement between theory and experiment is excellent. 

\begin{figure}[tbph]
\hspace{-.3cm}
\includegraphics[width=0.45\textwidth]{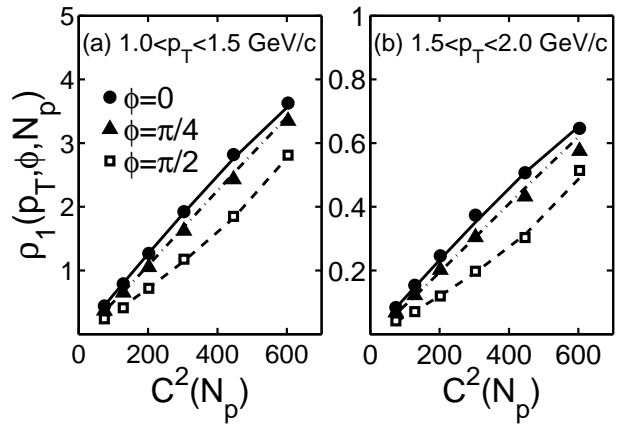}
\caption{Pion inclusive \dis\ $\rho_1(p_T,\phi,N_p)$ vs $C^2(N_p)$ at six points of $N_{\rm part}$  given in \cite{saf} for three ranges of $\phi: (0,15^{\circ}), (30^{\circ},45^{\circ})\ {\rm and}\ (75^{\circ},90^{\circ})$. (a) $1.0<p_T<1.5$ GeV/c, (b)  $1.5<p_T<2.0$ GeV/c. The lines are calculated values from Eqs.\ (\ref{31}) and (\ref{31a}).}
\end{figure}

Another way is to exhibit directly $R'_{AA}(p_T, \phi, N_p)$ vs $N_p$ in Fig.\ 8 and $R_{AA}(p_T, \phi, N_p)$ in Fig.\ 9. The calculated results based on Eq.\ (\ref{31}) are shown by solid, dash-dotted and dashed lines for $\phi=(0,15^{\circ}), (30^{\circ},45^{\circ})\ {\rm and}\ (75^{\circ},90^{\circ})$, respectively. They agree with the data essentially all within errors.  The spread for different values of $\phi$ is therefore to be interpreted as a consequence of  the physics of ridges. We now see that the flatness of the NMF $R_{AA}$ along the normal to the reaction plane, pointed out especially in \cite{saf}, is not caused by any extraordinary, single piece of physics.  It is due to the cancellation of the linearly rising behavior of $ R'_{AA}$ shown by the dashed lines in Fig.\ 8 and the decreasing behavior of $N_p/N_c$ in the rescaling factor $Z (N_p)/Z'(N_p)$ in Eqs.\ (\ref{4}) and  (\ref{5}).  The question is why  $R'_{AA}(\phi=\pi/2, N_p )$ rises with $N_p$.  At low $p_T$ path length is not an important issue.  The rising behavior is the consequence of a combination of factors that include the broadening of the tip of the ellipse, when $b$ is decreased, and the increasing likelihood for semihard partons directed at large $\phi_1$ to produce ridge particles with a narrow cone of angular correlation around $\phi_2 \approx \pi/2$.  That likelihood is quantified by $S(\phi, b)$, as shown by the dashed line in Fig.\ 3.

\begin{figure}[tbph]
\hspace{-.3cm}
\includegraphics[width=0.45\textwidth]{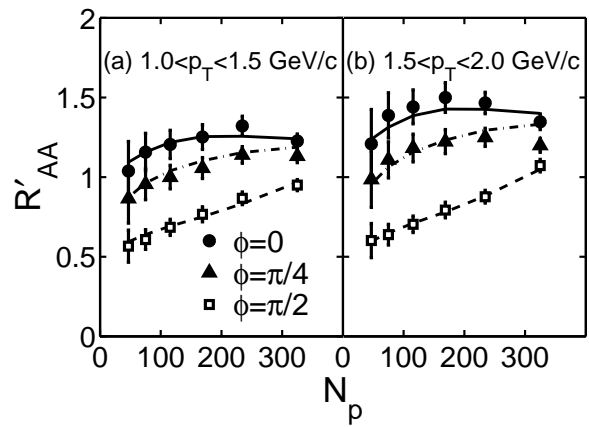}
\caption{Alternative $R'_{AA}(p_T,\phi,N_p)$ vs $N_p$ for $\phi$  and $p_T$ values as in the figure caption of Fig. 7. Data points are obtained from \cite{saf} by rescaling $R_{AA}(p_T,\phi,N_p)$. The lines are calculated results.}
\end{figure}

\begin{figure}[tbph]
\hspace{-.3cm}
\includegraphics[width=0.45\textwidth]{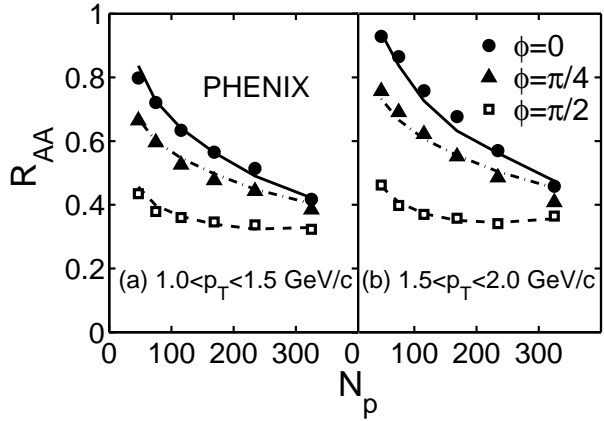}
\caption{Standard $R_{AA}(p_T,\phi,N_p)$ vs $N_p$. Data points are  from \cite{saf} as in Fig. 8. The lines are calculated results.}
\end{figure}
 
 \begin{figure}[tbph]
\hspace{-.3cm}
\includegraphics[width=0.45\textwidth]{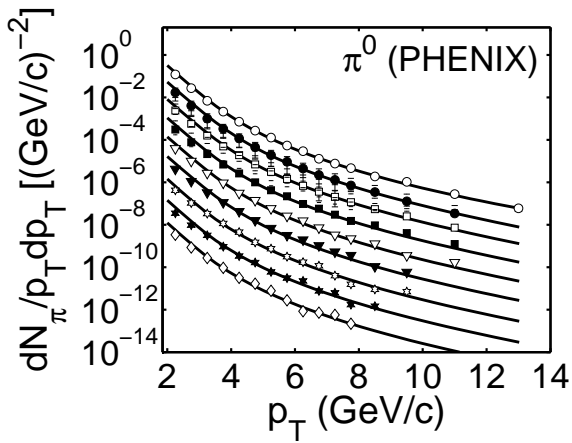}
\caption{Inclusive \dis s of $\pi^0$ for all centralities ranging from 0-10\% (top) to 80-90\% (bottom) in 10\% steps, each displaced by a factor of 0.2. The data are from \cite{saf}.}
\end{figure}

 In the above we have considered only pion production at low \pt, where TT recombination dominates. For $p_T>2$ GeV/c, the contributions from TS and SS recombination must be added, as have been calculated before for all centralities \cite{hy2}. But now with the new parametrization of $C(N_p)$ given in Eq.\ (\ref{32}) for TT recombination, we can revisit the single-particle \dis s at higher \pt,  averaged over all $\phi$, even though the TT component is not dominant.  The calculated result is shown in Fig.\ 10 for nine bins of centrality.  The agreement with data \cite{ssa} is evidently very good. How they depend on $\phi$ remains to be investigated.
 
 With this section we have completed the circuit and returned to the original question about the characteristics of the nuclear modification factor raised in Sec. I. By reproducing the data on its $\phi$ dependence at low $p_T$, we have demonstrated that all $\phi$ dependencies in ridges, elliptic flow and NMF are interconnected.

\section{CONCLUSION}

We have given a unified description of all the azimuthal dependencies of all observables on $\pi^0$ production at low \pt\ in heavy-ion collisions. They are: $R_{AA}(\phi,N_p), v_2(N_p)$ and ridge yield $Y^R(\phi_s,N_p)$ as functions of $N_p$, the number of participants. The main physics input is that the semihard scattering near the surface drives the azimuthal anisotropy on the one hand and the production of ridge particles on the other with or without trigger. The geometrical factor, $S(\phi,b)$, that makes precise the bridge between the two aspects of the problem follows from a study of the correlation between the $\phi$ directions of the trigger and ridge particles. The main understanding achieved in this picture is that the single-particle \dis\ $dN_{AA}^{\pi}/p_Tdp_Td\phi$ has two components: one is the $\phi$-independent bulk $B(p_T,N_p)$, different from the conventional bulk that is $\phi$-dependent, and the other is the ridge component $R(p_T,\phi,N_p)$ that carries all the $\phi$ dependence. It is the latter that can be tested by correlation experiment using triggers.

A number of parameters are  involved in this analysis, but most of them are adopted from previous studies. The correlation width $\sigma$ is an important input, and its value was determined in Ref.\ \cite{ch} in connection with the  study of ridgeology. The values of $T$ and $T'$ are from Refs.\ \cite{jp} and \cite{hy}, and that of $\delta r$ is an approximation of what was found relevant in Ref.\ \cite{ch}. The parametrization of $C(N_p)$ is an improvement of that found in Ref.\ \cite{yt} and is a description of the dependence on centrality, not on $\phi$. The one parameter that we have adjusted in this paper to fit the normalization of $v_2(p_T,b)$ is $a$, which appears in Eq.\ (\ref{20}) and affects the magnitude of the ridge. The basic properties of the $\phi$ dependence that we can calculate, such as $S(\phi,b)$ and $Y^R(\phi_s,b)$ do not depend on $a$, although $R_{AA}(p_T,\phi,b)$ depends on it indirectly.

The concern in this paper has been exclusively about low-\pt\ pions at $p_T<2$ GeV/c. Conventionally, such particles are regarded as products of soft processes. However, our view is that some of them can be due to semihard processes that can copiously produce semihard partons whose footprints are the ridge particles at low \pt. Triggers select a subset of events for the study of correlations, but without triggers the effect of the semihard partons must nevertheless be taken into account in the calculation of inclusive \dis s at low \pt.

The production of proton has not been considered, either in this paper or  in the experimental study of the $\phi$ dependence of $R_{AA}$ \cite{saf}. Theoretically, large baryon/meson ratio has been a signature of the recombination model \cite{rev, qgp}. Since the same hadronization mechanism applies to both inclusive and ridge production, we expect that $R_{AA}$ for proton will have the same characteristics as that for pion. There are some differences in the \pt\ dependence because of the difference in the recombination functions, but the physics connecting $R_{AA}$ and $Y^R$ is the same. That is, at a fixed centrality $S(\phi,b)$ exhibits the range of variation when  $\phi$  is varied  from 0 to $\pi/2$, which in turn describes the ranges of variation of $Y^R(\phi_s)$ and $R_{AA}(\phi)$, as shown in Figs.\ 5 and 9, respectively. The behavior for proton production should be similar. This is a good testing ground for any exotic model that attempts to explain the so-called ``baryon anomaly".

At intermediate \pt\ above 2 GeV/c the dominance of TS recombination changes the physics considered here. Ridge is a manifestation of the enhanced thermal partons, and the peak sitting
above the ridge is localized in $\Delta\eta$ and is more closely associated with the hard parton through the intermediary shower partons \cite{qgp}. Correlation in a jet peak is different from that in the ridge, so the geometrical effect is also different. Although the data on $R_{AA}(p_T,\Delta\phi,N_p)$ vs $\Delta\phi$ in \cite{saf} do not show significant difference between $1.0<p_T<1.5$ and $4.0<p_T<5.0$ GeV/c sectors, their $N_p$ dependencies are quite different. For example, $R_{AA}(\Delta\phi\approx \pi/2)$ is not constant in $N_p$ for $4<p_T<5$ GeV/c, in contrast to the behavior at $1.0<p_T<1.5$ GeV/c. It is clear that the physics at intermediate \pt\ is very different and requires a fresh approach that is not a simple extrapolation from the study described here. However, the TT component at higher $p_T$, though smaller, can be reliably calculated only if the formalism developed here is used for the description of the $\phi$ dependence. Such a study has been carried out and a remarkable scaling behavior has been found in the dual dependence on $\phi$ and centrality \cite{hy3}, a property that is contained in the PHENIX data \cite{saf}.

\section*{Acknowledgment}
We are grateful to Chunbin Yang for helpful discussions. This work was supported,  in part,  by the U.\ S.\ Department of Energy under Grant No. DE-FG02-96ER40972 and by the National Natural Science Foundation of China under Grant No. 10635020 and 10775057 and by the Ministry of Education of China under project IRT0624.


\end{document}